\documentclass[11pt,a4paper,headrule]{article}
\usepackage[body={80mm, 210mm},left=2.5cm,right=2.5cm,
headheight=2cm, headsep=5cm,voffset=.5cm,hoffset=0pt]{geometry}
\usepackage{amsfonts,amscd,amsmath,enumerate,verbatim,calc}
\usepackage{amssymb}
\usepackage{derivative}
\usepackage{mathtools}
\usepackage{amsthm}
\usepackage{epsfig}
\usepackage{graphics}
\usepackage{color}
\usepackage{graphics}
\usepackage{graphicx}
\usepackage{lineno}
\usepackage{bigints}
\usepackage{caption}
\usepackage{subcaption}
\usepackage{xfrac}  
\usepackage{multirow}
\usepackage{float}
\usepackage{tablefootnote}

\newcommand*\diff{\mathop{}\!\mathrm{d}}

\begin{document}

		\title{\bf{Refining Boundary Value Problems in Non-local Micropolar Mechanics}}
	\author{Manasa Bhat and Santanu Manna$^*$\\[4pt]
		Department of Mathematics, Indian Institute of Technology Indore,\\ Simrol, Indore, Madhya Pradesh 453552, India}
	\date{}
	\maketitle
	\let\thefootnote\relax\footnotetext{\hspace{-.65cm}  $^{*}$Corresponding Author.\\{E-mail addresses}:  phd2101141004$@$iiti.ac.in and santanu$@$iiti.ac.in$^{*}$}
    \textbf{Abstract:} 
    This research explores refined boundary conditions for a traction-free surface in a non-local micropolar half-space, combining non-local and micropolar elasticity effects to study Rayleigh wave propagation in an isotropic, homogeneous medium. This study revisits the solution for Rayleigh waves obtained within the framework of Eringen's non-local differential model. It highlights that the equivalence between the non-local differential and integral formulations breaks down for a micropolar half-space and can only be restored under specific additional boundary conditions. For mathematical tractability, equivalence is assumed for a defined subset of stresses. Asymptotic analysis is further employed to capture the effects of the boundary layer within the non-local micropolar half-space. This technique finally derives the refined boundary conditions for micropolar media.   

 \section{Introduction}
    \par Non-local elasticity emerges as a robust framework ideally suited for examining the characteristics of solids where long-range atomic interactions exert a substantial influence. Researchers aiming to engineer micro/nanostructures have shown considerable interest in exploring these non-local elastic models. These models overcome the limitations of classical elasticity theory by focusing on the long-range interactions between the material particles at small scales, such as molecules or atoms. Unlike classical elasticity, non-local models focus on how strains far away influence the stress at a specific point. These models, like those proposed in references \cite{Eringen1983, Eringen1972a, Eringen1977a}, assume the non-local nature of stress, where the stress at a point depends not only on the local strain but also on the strain field distributed across the entire material body. This influence of distant strains is captured through integral equations in the constitutive relations. The groundwork for non-local elasticity is attributed to A.C. Eringen \cite{Eringen1987}. He proposed a strain-driven integral formulation, where stress is a convolution integral of the elastic strain field and a kernel function incorporating an internal characteristic length. It is important to note the contributions of Kroner \cite{Kroner1967}, Eringen et al. \cite{Eringen1977b}, Eringen and Kim \cite{Eringen1974}, Eringen and Edelen \cite{Eringen1972b}, Edelen et al. \cite{Edelen1971} in establishing the fundamental principles of non-local elasticity during its formative years.
    \par Nevertheless, incorporating long-range forces alone proves inadequate for capturing the intricacies of the dynamical responses of a material. To address this shortcoming, advanced continuum theories have been developed. One such theory is the micropolar elasticity theory, which specifically focuses on the inherent internal structure of a material \cite{Eringen1966}. This theory acknowledges that materials are not just continuous media but have a microstructure that influences their behavior. It considers microrotations of tiny material elements alongside the overall deformation. By virtue of their inherent microstructure, micropolar materials \cite{Eringen1999} exhibit additional degrees of freedom that characterize their local rotations. Deformation in these solids encompasses not only displacement but also microrotation of their constituent elements. Furthermore, couple stress is introduced alongside classical stress within the constitutive framework to account for the internal rotations. Integrating the concepts of non-local elasticity with micropolar elasticity holds promise for a more comprehensive understanding of complex material behavior. Several researchers have undertaken a detailed examination of non-local and micropolar effects in various media, investigating them both independently \cite{Bhat2023, Kumar2023, Bhat2024} and collectively \cite{Kumar2020, Vinh2023, Tung2021}.
    \par Despite extensive research, a key challenge remains in applying micropolar non-local elasticity theory is to account for boundaries. Classical non-local constitutive relations in an infinite media use integral operators that integrate the strain field over the entire domain. However, in a half-space, the appropriate integration interval for these operators depends on the distance from a point to the material boundary \cite{Eringen1983}.  This dependence on the global strain field induces the formation of a localized boundary layer near the surface, characterized by non-homogeneous distributions of stress and strain fields. Boundary layer theory and asymptotic analysis are considered the most valuable tools in geomechanics for investigating the impact of boundary layers on the overall complex non-local behavior of geological materials. By applying these theories, one can develop effective models and solutions to address challenges related to soil-structure interaction \cite{Kim2016}, landslide prediction, and other geotechnical applications. Integrating boundary layer concepts and asymptotic techniques enhances our understanding of the mechanical response of earth materials and supports the development of safer and more sustainable engineering practices in geomechanics. Furthermore, various authors \cite{Abdollahi2014, Romano2018} have emphasized employing a boundary layer method to address the numerical challenges associated with Eringen's non-local elasticity model.
    \par The application of linear non-local elasticity theory for homogeneous, isotropic materials leads to a system of integropartial differential equations. While these equations generally present significant challenges in terms of solution techniques, assuming a specific class of non-local kernels allows for their reduction to a set of singular partial differential equations \cite{Eringen1983}. The equivalence between differential and integral formulations of non-local elasticity is well-established for unbounded media. However, for bounded media, the presence of boundary layers introduces complexities, raising questions about the equivalence of the two formulations. Researchers have proposed specific approaches to address the challenges of equivalence in bounded media. Chebakov et al. \cite{Chebakov2016, Chebakov2017} derived appropriate boundary conditions for non-local elastic half-spaces and plates under the assumption of a small internal length scale compared to the characteristic wavelength. Additionally, Kaplunov et al. \cite{Kaplunov2022, Kaplunov2023} and Sahin et al. \cite{Sahin2024, Sahin2023} employed asymptotic analysis to derive effective boundary conditions expressed in terms of local stresses. These works highlight the need for corrections to conventional boundary conditions due to the influence of boundary layers in bounded non-local elastic media. To the best of the authors' knowledge, a critical gap that remains in the current literature is the application of asymptotic analysis to establish refined boundary conditions for non-local micropolar elastic half-spaces.
    \par This paper focuses on the derivation of effective boundary conditions on the free surface in a non-local micropolar elastic half-space. The problem is motivated by the limitations of the classical approach, where stress depends solely on local strain. Here, the influence of the entire strain field is considered, leading to non-homogeneous behavior near the surface (boundary layer effect). Section 2 establishes the governing equations for this material, incorporating a specific non-local kernel taking the form of the Bessel function. Section 3 reveals that the equivalence between standard formulations of non-local elasticity breaks down in this context.  To address the boundary layer, Section 4 employs asymptotic analysis to solve the resulting equations. Finally, Section 5 presents the derived refined boundary conditions.

 \section{Equations for non-local micropolar elasticity}
 Here, we will write the constitutive stress-strain relations and equations of motion for the surface waves propagating in a non-local micropolar elastic medium. This will serve as the fundamental basis for our study.\\
 A micropolar elastic solid exhibits both translational and rotational motions, which give rise to force stresses and couple stresses within the medium. The constitutive relations that relate these local stresses with the strain tensor $(\epsilon)$ and curvature tensor $(\Gamma)$ are given as \cite{Eringen1999},
     \begin{eqnarray}
	\renewcommand{\arraystretch}{1.1}
	\left. \begin{array}{ll}
		\sigma_{mn}=\Lambda\,\varepsilon_{pp}\,\delta_{mn}+(\mu+\kappa)\,\varepsilon_{mn}+\mu\,\varepsilon_{nm},\\
		\Pi_{mn}=\alpha\,\Gamma_{pp}\,\delta_{mn}+\beta\,\Gamma_{mn}+\gamma\,\Gamma_{nm},
	\end{array}
	\right\} \, m,n=1,2,3. \label{1}
    \end{eqnarray}
Here $\kappa, \alpha,\beta, \gamma$ are the micropolar constants; $\Lambda, \mu$ are the elastic constants; $\delta_{mn}$ denotes the well-known Kronecker delta symbol. 
Moreover, the relation between the curvature tensor and strain tensor with the displacement components and rotational components are given as,
\begin{eqnarray}
	\renewcommand{\arraystretch}{1.1}
	\left. \begin{array}{ll}
		\varepsilon_{mn}=u_{n,m}-\epsilon_{mnp}\Phi_p,\\
		\Gamma_{mn}=\Phi_{m,n},
	\end{array}
	\right\} \, m,n=1,2,3. \label{2}
\end{eqnarray}
where $u_{n}$ and $\Phi_n,~ n=1,2,3$ are the displacement components and microrotation vector components of the surface wave, respectively.\\
Given that the model is also explored within the framework of non-local elasticity theory, it is essential to formulate the expressions for the non-local force stresses and couple stresses.\\
According to Eringen's theory, the non-local stresses $(\tau_{mn})$ and local stresses $(\sigma_{mn})$ are related to each other through a singular kernel $\beta$ as follows,
\begin{equation}
	\tau_{mn}(\textbf{x})=\int_{V}\beta(\left|\textbf{x}-\textbf{x}'\right|)\,\sigma_{mn}(\textbf{x}')\,\text{d}V(\textbf{x}'). \label{3}
\end{equation}
Here, $\beta$ represents a non-locality kernel that depends on the material properties of the medium, and $V$ denotes the volume of the region over which the deformation has occurred.\\
Consider the two-dimensional singular non-local kernel as provided in Eringen's theory \cite{Eringen1983}, 
\begin{equation}
    \alpha\left(|\textbf{x}'-\textbf{x}|,\tau\right)=\frac{1}{2\pi \mathfrak{a}^2}\,K_0\left(\frac{\sqrt{\left(\textbf{x}'-\textbf{x}\right)\,.\,\left(\textbf{x}'-\textbf{x}\right)}}{\mathfrak{a}}\right) \label{4}
\end{equation}
where $\mathfrak{a}$ is the non-locality parameter associated with the micropolar medium.\\
Consequently, the relationship between conventional local stresses and non-local stresses can be established through the integral formulation \cite{Kaplunov2023} as follows:
\begin{eqnarray}
\renewcommand{\arraystretch}{1.8}
\left. \begin{array}{ll}
    \tau_{mn}=\frac{1}{2\pi \mathfrak{a}^2}\bigint_{0}^{\infty}\bigint_{-\infty}^{\infty}K_0\left(\frac{\sqrt{(x-x')^2+(z-z')^2}}{\mathfrak{a}}\right)\sigma_{mn}\left(x',z'\right)\,\diff x'\, \diff z' \\
    \mathfrak{M}_{mn}=\frac{1}{2\pi \mathfrak{a}^2}\bigint_{0}^{\infty}\bigint_{-\infty}^{\infty}K_0\left(\frac{\sqrt{(x-x')^2+(z-z')^2}}{\mathfrak{a}}\right)\Pi_{mn}\left(x',z'\right)\,\diff x'\, \diff z'
    \end{array}
\right\} \label{5}
\end{eqnarray}
Additionally, utilizing the same two-dimensional kernel as specified in Eq. \eqref{4}, the differential model proposed by Eringen \cite{Eringen1983} relates the local and non-local stresses as,
\begin{eqnarray}
    \renewcommand{\arraystretch}{1.1}
    \left. \begin{array}{lr}
         \left(1-\mathfrak{a}^2\nabla^2\right)\tau_{mn}=\sigma_{mn}\\
         \left(1-\mathfrak{a}^2\nabla^2\right)\mathfrak{M}_{mn}=\Pi_{mn}
     \end{array}
    \right\} \label{6}
\end{eqnarray}
Introduce the dimensionless variables,
\begin{equation*}
    \chi=\frac{x}{\lambda},~~\eta=\frac{z}{\lambda}, 
\end{equation*}
and 
\begin{equation*}
    ~~\epsilon=\frac{\mathfrak{a}}{\lambda}<<1
\end{equation*}
is the small dimensionless parameter associated with non-locality in the micropolar medium.\\
Following Kaplunov et al. \cite{Kaplunov2023}, and with the assumption of slow variation of local stresses $\sigma_{mn}$ along $x-$direction, we approximate Eqs. \eqref{4} as,
\begin{eqnarray}
    \tau_{mn}&\approx&\frac{1}{2\epsilon}\int_{0}^{\infty}\left[1+\frac{\epsilon^2}{2}\left(1+\frac{|\eta'-\eta|}{\epsilon}\right)\frac{\partial^2}{\partial \chi^2}\right]\sigma_{mn}(\chi,\eta')\,\text{exp}\left(-\frac{|\eta'-\eta|}{\epsilon}\right)\,\diff \eta' \label{7}\\
    \mathfrak{M}_{mn}&\approx&\frac{1}{2\epsilon}\int_{0}^{\infty}\left[1+\frac{\epsilon^2}{2}\left(1+\frac{|\eta'-\eta|}{\epsilon}\right)\frac{\partial^2}{\partial \chi^2}\right]\Pi_{mn}(\chi,\eta')\,\text{exp}\left(-\frac{|\eta'-\eta|}{\epsilon}\right)\,\diff \eta' \label{8}
\end{eqnarray}
 Further, the governing equations for Rayleigh waves propagating in a non-local micropolar elastic solid (without body forces) are given by,
 \begin{eqnarray}
 \renewcommand{\arraystretch}{1.1}
    \left. \begin{array}{lr}
        \tau_{1n,x}+\tau_{3n,z}-\rho\,u_{n,tt}=0,\\
       \mathfrak{M}_{12,x}+\mathfrak{M}_{32,z}+\tau_{31}-\tau_{13}-\rho \, j\, \Phi_{2,tt}=0
    \end{array}
    \right\}~~n=1, 3 \label{9}
 \end{eqnarray}
 Using the differential formulation of non-local stresses as in Eq. \eqref{6}, the equation of motion in the form of local stresses can be written as,
 \begin{eqnarray}
    \renewcommand{\arraystretch}{1.1}
    \left. \begin{array}{lr}
        \sigma_{1n,x}+\sigma_{3n,z}-\rho\,\left(1-\mathfrak{a}^2\nabla^2\right)\,u_{n,tt}=0,\\
        \Pi_{12,x}+\Pi_{32,z}+\sigma_{31}-\sigma_{13}-\rho \, j\, \left(1-\mathfrak{a}^2\nabla^2\right)\,\Phi_{2,tt}=0
    \end{array}
    \right\}~~n=1,3 \label{10}
 \end{eqnarray}
    To decouple the system in Eq. \eqref{10}, we shall apply the method of potentials in which the displacement components are decomposed into the sum of a scalar function $\phi$ and a vector potential $\Psi=(\psi_1,\psi,\psi_3)$ as,
    \begin{equation}
        u_1 (x,z,t)=\phi_{,x}-\psi_{,z}, ~~u_3(x,z,t)=\phi_{,z}+\psi_{,x} \label{11}
    \end{equation}
    This reduces the governing coupled equations of motion given in Eq. \eqref{10} as,
\begin{eqnarray}
	\renewcommand{\arraystretch}{1.1}
	\left. \begin{array}{rr}
		\left(\Lambda+2\,\mu+\kappa\right)\nabla^2\phi-\rho\left(1-\mathfrak{a}^2\nabla^2\right)\,\phi_{,tt}=0,\\
		\left(\mu+\kappa\right)\nabla^2\psi+\kappa\,\Phi_2-\rho \left(1-\mathfrak{a}^2\nabla^2\right)\,\psi_{,tt}=0,\\
		\gamma \nabla^2\Phi_2-\kappa \nabla^2\psi-2\,\kappa\,\Phi_2-\rho 	j\left(1-\mathfrak{a}^2\nabla^2\right)\Phi_{2,tt}=0.
	\end{array}
	\right\} \label{12}
\end{eqnarray}
Let us define certain velocity parameters for the micropolar elastic media as
\begin{equation*}
    c_1=\sqrt{\frac{\Lambda+2\mu+\kappa}{\rho}}, ~~~ c_2=\sqrt{\frac{\mu+\kappa}{\rho}},~~~ c_3=\sqrt{\frac{\kappa}{\rho}}, ~~~c_4=\sqrt{\frac{\gamma}{\rho j}}.
\end{equation*}
and assume that the solutions to the system of equations to be propagating in a time-harmonic form as,
\begin{equation}
	\{\phi,\psi,\Phi_2\}=\{P, Y, Z\}\,e^{-krz} e^{i\left(k x-\omega t\right)}, \label{13}
\end{equation}
 Substituting Eq. \eqref{13} in decoupled system of equations of Eq. \eqref{12}, we get,
 \begin{eqnarray}
	\renewcommand{\arraystretch}{1.1}
	\left. \begin{array}{lr}
        \phi(x,z,t)=P \,e^{-kr_1z} e^{i\left(k x-\omega t\right)},\\
		\psi(x,z,t)=\left[Q\,e^{-kr_2z}+R\,e^{-kr_3z}\right]\,e^{i\left(k x-\omega t\right)},\\
        \Phi_2(x,z,t)=s\,k^2\,R\,e^{-kr_3z}\,e^{i\left(k x-\omega t\right)},
	\end{array}
	\right\} \label{14}
\end{eqnarray}
where
\begin{eqnarray*}
   &&r_1^2=1-\frac{v^2}{c_1^2-\epsilon ^2 v^2 },~~
	r_2^2=1-\frac{v^2}{c_2^2-\epsilon ^2v^2 },~~\\
    && r_3^2=1-\frac{v^2}{c_4^2- \epsilon ^2v^2} \left(1-\frac{2 c_3^2}{j \omega ^2}\right),~~
    s=\frac{v^2}{c_3^2} \left[1-\frac{c_2^2-\epsilon ^2v^2 }{{c_4^2-\epsilon ^2 v^2}} \left(1-\frac{2\,c_3^2}{j \omega ^2}\right)\right],
\end{eqnarray*}
are such that Re$(r_i)>0$ for $i=1,2,3$ to ensure the exponential decay of the waves from the surface.\\
The traction-free boundary conditions at the surface $z=0$ indicate the vanishing of force-stresses and couple-stresses at the surface, i.e., at $z=0$,
\begin{equation}
    \tau_{31}=0, ~~~\tau_{33}=0, ~~~\mathfrak{M}_{32}=0 \label{15}
\end{equation}
\section{Failure of equivalence}
This section aims to verify the equivalence between the integral and differential formulations of Eringen's non-local elasticity theory. We will utilize the differential form of the non-local model in the equation of motion, while employing the integral formulation for the boundary conditions, and subsequently assess their equivalence. For our convenience, we will omit the time-harmonic terms for further studies.\\
Now, the non-local stresses can be explicitly written from Eqs. \eqref{7} and \eqref{10} using Eq. \eqref{1} as,
\begingroup\makeatletter\def\f@size{9.5}\check@mathfonts 
\begin{eqnarray*}
    \tau_{11}&=&k^2\left[-\left(\Lambda +2 \mu +\kappa-\Lambda\, r_1^2\right)\,P\,I_1+i\,r_2\,(2 \mu+\kappa )\,Q\,I_2+i\,r_3\,(2 \mu+\kappa)\,R\,I_3\right]\\
    \tau_{13}&=&k^2\,\left[-i\,r_1\,(2 \mu+\kappa)\,P\,I_1-\left(\mu+\kappa+\mu\,r_2^2\right)\,Q\,I_2+\left(s\,\kappa -(\mu+\kappa)- \mu\, r_3^2\right)\,R\,I_3 \right]\\
    \tau_{31}&=&k^2\,\left[-i\,r_1\,(2 \mu+\kappa)\,P\,I_1-\left(\mu + (\mu +\kappa)\,r_2^2\right)\,Q\,I_2-\left(s\,\kappa+ \mu +  (\mu +\kappa)\, r_3^2\right)\,R\,I_3\right]\\
    \tau_{33}&=&k^2\,\left[\left((\Lambda+2 \mu+\kappa)\,r_1^2-\Lambda \right)\, P\,I_1-i\,r_2\,(2 \mu +\kappa)\,Q\,I_2-i\,r_3\,(2 \mu +\kappa)\,R\,I_3\right]\\
    \mathfrak{M}_{12}&=&i\,s\,k^3\,\gamma\,R\,I_3\\
    \mathfrak{M}_{32}&=&-s\,k^3\,\gamma\,r_3\,R\,I_3
\end{eqnarray*}
\endgroup
where 
\begingroup\makeatletter\def\f@size{9}\check@mathfonts 
\begin{equation}
    I_i=\frac{1}{2 \mathfrak{a}}\int_{0}^{\infty}\left[1+\frac{\mathfrak{a}^2}{2}\left(1+\frac{|x'-x|}{\mathfrak{a}}\right)\frac{\partial^2}{\partial x^2}\right]\,\exp{\left(ikx-kr_iz\right)}\,\exp{\left(-\frac{|z'-z|}{\mathfrak{a}}\right)}\,\diff z',~~~i=1,2,3 \label{16}
\end{equation}
\endgroup
On splitting $I_i$ in Eq. \eqref{16} over the interval from $0$ to $z$ and from $z$ to $z'$ following the approach of Kaplunov et al. \cite{Kaplunov2023}, and subsequently simplifying, we obtain,
\begin{equation}
    I_i=\left[1+\epsilon^2\left(r_{i0}^2-1\right)\right]\,\text{exp}\left(-k r_i z\right)-\frac{1}{2}\left[1+\epsilon\,r_{i0}+\epsilon^2\left(r_{i0}^2-1-\frac{ k z}{2 \epsilon}\right)\right]\,\text{exp}\left(-\frac{ k z}{\epsilon}\right) \label{17}
\end{equation}
The coefficients of $\text{exp}\left(\frac{-kz} {\epsilon}\right)$ in Eq. \eqref{17} are linked with the boundary layer formed as a result of the non-local elasticity effects in the medium.\\
Substituting in Eq. \eqref{15}, a system of equations in $P, Q, R$ is obtained. This system is then solved for the non-trivial solution to obtain the dispersion relations for Rayleigh waves within an error of $O\left(\epsilon^2\right)$ as,
\begin{eqnarray}
   (1+d)^2\,r_{10}\,r_{20}-\left(r_{20}^2+d\right)^2&=&0 \label{18}\\
   r_{30}&=&0 \label{19}
\end{eqnarray}
with $d=\frac{\mu}{\mu+\kappa}$. Also, $r_{10},r_{20}$ and $r_{30}$ are the leading order Taylor series approximations of $r_1,r_2$ and $r_3$, respectively within an error of $O\left(\epsilon^2\right)$ and are given by,
\begin{equation*}
    r_{10}^2=1-\frac{v^2}{c_1^2}, ~~r_{20}^2=1-\frac{v^2}{c_2^2}, ~~ r_{30}^2=1-\frac{v^2}{c_4^2}\left(1-\frac{2 c_3^2}{j \omega^2}\right)
\end{equation*}
Eqs. \eqref{18} and \eqref{19} represent the dispersion relations consisting of two modes of Rayleigh waves, one of which is entirely due to the micropolarity in the medium. Also, these two modes do not coexist at the same period. In other words, the relations \eqref{18} and \eqref{19} cannot be zero simultaneously.\\
For the mode corresponding to Eq. \eqref{18} and from the boundary conditions at $z=0$, the arbitrary constants $P$, $Q$ and $R$ are related by,
\begin{eqnarray*} 
    P=\frac{i\,\left(r_{20}^2+d\right) \left(1+\left(r_{10}-r_{20}\right) \left(\epsilon -\epsilon ^2\,r_{20} \right)\right)}{(1+d)\,r_{10}}\,Q,~~~
    R=0 
\end{eqnarray*}
Substituting these values of $P, Q, R$ into the equation of motion given in \eqref{9} for $m=1$ and analyzing the leading order term that is associated with the boundary layer near the vicinity of the surface, we observe that,
\begin{eqnarray}
    \frac{k^3}{2 (1+d)^2\,r_{10}}\left[(1+d)^2\,r_{10}^2-2\,r_{20}^2 \left(d+r_{20}^2-1\right)-(1+d^2)\right]\exp{\left(-\frac{k z}{\epsilon}\right)}\neq 0 \label{20}
\end{eqnarray}
Similarly, for the micropolar mode corresponding to Eq. \eqref{19}, the relation between the arbitrary constants is given as,
\begin{eqnarray*}
   P&=&\frac{i\, (1+d)\, r_{20}}{r_{20}^2+d}\,\left[1+\left(r_{10}-r_{20}\right)\left(\epsilon-\epsilon^2\, r_{20}\right)\right]\,Q\\
   R&=&\frac{\left(r_{20}^2+d\right)^2-(1+d)^2\,r_{10}\,r_{20}}{r_{20}^2+d}\,\left[1+\left(r_{30}-r_{20}\right)\left(\epsilon-\epsilon^2\, r_{20}\right)\right]\,Q,
\end{eqnarray*}
Substituting the same into the equation of motion \eqref{9} at $m=1$, we have at the leading order,
\begin{eqnarray}
    \frac{k^3}{r_{20}^2+d}\left[(1+d)^2\,r_{10}\,r_{20}-\left(r_{20}^2+d\right)^2\right]\,\exp{\left(\frac{-kz}{\epsilon}\right)}\neq0 \label{21}
\end{eqnarray}
which is true as this mode does not coexist with the mode corresponding to Eq. \eqref{18}.\\
Eqs. \eqref{20} and \eqref{21} indicate that the solution for the Rayleigh wave obtained from a differential non-local model does not comply with the original governing equations derived under the framework of the integral non-local elastic model. This suggests the failure of the equivalence between the differential and integral formulations of the non-local model in case of a non-local micropolar half-space.\\~\\
Upon substituting the differential form of the local stresses provided in Eq. \eqref{6} into Eqs. \eqref{7} and \eqref{8}, which represents the integral formulation of the non-local stresses at the boundary, we get,
\begingroup\makeatletter\def\f@size{9.5}\check@mathfonts 
\begin{eqnarray}
    \tau_{mn}\Bigr|_{\eta=0}&=&\frac{1}{2\epsilon}\int_{0}^{\infty}\left[1+\frac{\epsilon^2}{2}\left(1+\frac{\eta'}{\epsilon}\right)\frac{\partial^2}{\partial \chi^2}\right]\left[1-\epsilon^2\left(\frac{\partial^2}{\partial \chi^2}+\frac{\partial^2}{\partial \eta'^2}\right)\right]\, \tau_{mn}\left(\chi, \eta'\right)\,e^{-\frac{\eta'}{\epsilon}}\,\diff \eta' \label{22}\\
    \mathfrak{M}_{mn}\Bigr|_{\eta=0}&=&\frac{1}{2\epsilon}\int_{0}^{\infty}\left[1+\frac{\epsilon^2}{2}\left(1+\frac{\eta'}{\epsilon}\right)\frac{\partial^2}{\partial \chi^2}\right]\left[1-\epsilon^2\left(\frac{\partial^2}{\partial \chi^2}+\frac{\partial^2}{\partial \eta'^2}\right)\right]\, \mathfrak{M}_{mn}\left(\chi, \eta'\right)\,e^{-\frac{\eta'}{\epsilon}}\,\diff \eta' \label{23}
\end{eqnarray}
\endgroup
This is further simplified by ignoring the higher order terms of $O(\epsilon^4)$ to get,
\begin{eqnarray}
        \tau_{mn}\Bigr|_{\eta=0}&=&\frac{1}{2}\left[\tau_{mn}\Bigr|_{\eta=0}+\epsilon\,\frac{\partial \tau_{mn}}{\partial \eta}\Biggr|_{\eta=0}+\frac{\epsilon^3}{2}\frac{\partial^3\tau_{mn}}{\partial \chi^2\,\partial \eta}\Biggr|_{\eta=0}\right] \label{24}\\
        \mathfrak{M}_{mn}\Bigr|_{\eta=0}&=&\frac{1}{2}\left[\mathfrak{M}_{mn}\Bigr|_{\eta=0}+\epsilon\,\frac{\partial \mathfrak{M}_{mn}}{\partial \eta}\Biggr|_{\eta=0}+\frac{\epsilon^3}{2}\frac{\partial^3\mathfrak{M}_{mn}}{\partial \chi^2\,\partial \eta}\Biggr|_{\eta=0}\right] \label{25}
\end{eqnarray}
Further rearrangement gives a set of conditions,
\begin{eqnarray}
    \left[1- \epsilon\,\frac{\partial}{\partial \eta}-\frac{\epsilon^3}{2}\,\frac{\partial^3}{\partial \chi^2\,\partial \eta}\right]\,\tau_{mn}\Biggr|_{\eta=0}&=&0 \label{26}\\
     \left[1- \epsilon\,\frac{\partial}{\partial \eta}-\frac{\epsilon^3}{2}\,\frac{\partial^3}{\partial \chi^2\,\partial \eta}\right]\,\mathfrak{M}_{mn}\Biggr|_{\eta=0}&=&0  \label{27}
\end{eqnarray}
for which the equivalence between the differential and integral non-local model is possible.\\
However, for the equations of motion described in Eq. \eqref{9}, the boundary conditions derived from Eqs. \eqref{26} and \eqref{27}, along with the already prescribed boundary conditions in Eq. \eqref{15}, result in an ill-posed problem. Thus, we conclude that not all the conditions in Eqs. \eqref{26} and \eqref{27} can be satisfied at the boundary. \\
Instead, we redefine the proposed problem by considering the singularly perturbed differential equations as
\begin{eqnarray}
&&\hspace{1cm}    \renewcommand{\arraystretch}{1.1}
    \left. \begin{array}{lr}
        \tau_{1n,x}+\tau_{3n,z}-\rho\,u_{n,tt}=0,~~ n=1,3\\
        \mathfrak{M}_{12,x}+\mathfrak{M}_{32,z}+\tau_{31}-\tau_{13}-\rho \, j\, \Phi_{2,tt}=0
    \end{array}
	\right\} \label{28}\\
 &&\text{and}~~~
    \renewcommand{\arraystretch}{1.1}
	\left. \begin{array}{lr}
        \mathfrak{a}^2\left(\tau_{mn,xx}+\tau_{mn,zz}\right)-\tau_{mn}=-\sigma_{mn},\\
        \mathfrak{a}^2\left(\mathfrak{M}_{m2,xx}+\mathfrak{M}_{m2,zz}\right)-\mathfrak{M}_{m2}=-\Pi_{m2},
    \end{array}
	\right\} ~~m=1,3 \label{29}
\end{eqnarray}
with the boundary conditions prescribed on the surface $z=0$ as,
\begin{gather}
    \tau_{31}\bigr|_{z=0}=0, ~~~\tau_{33}\bigr|_{z=0}=0,~~~ \mathfrak{M}_{32}\bigr|_{z=0}=0 \label{30}\\
    \left[1- \mathfrak{a}\,\frac{\partial}{\partial z}-\frac{\mathfrak{a}^3}{2}\,\frac{\partial^3}{\partial x^2\,\partial z}\right]\,\tau_{11}\Biggr|_{z=0}=0,~~
     \left[1- \mathfrak{a}\,\frac{\partial}{\partial z}-\frac{\mathfrak{a}^3}{2}\,\frac{\partial^3}{\partial x^2\,\partial z}\right]\,\mathfrak{M}_{12}\Biggr|_{z=0}=0 \label{31}
\end{gather}
This suggests that by employing the singularly perturbed differential model as defined above, an equivalence between the differential and integral formulation can be obtained for the non-local force stress $t_{11}$ and couple stress $\mathfrak{M}_{12}$.
\section{Asymptotic analysis}
Introduce the dimensionless parameter,
\begin{equation*}
    \epsilon=\frac{\mathfrak{a}}{\lambda},
\end{equation*}
to aid in obtaining the solutions for the singularly perturbed differential equations via asymptotic analysis.
The presence of this small parameter $\epsilon$ leads to two distinct scales of behavior in the differential equation. To interprete this behavior, we introduce the following slow and fast dimensionless variables along the length scale as,
\begin{equation*}
    \eta_s=\frac{z}{\lambda}, ~~\eta_f=\frac{\eta}{\mathfrak{a}}.
\end{equation*}
Here, $\eta_s$ represents the slow behavior of the system, governed by simpler dynamics, while $\eta_f$ captures the rapid changes occurring in the boundary layer of the system due to the presence of $\epsilon$.\\
Further, define the following dimensionless parameters,
\begin{gather*}
    \chi=\frac{x}{\lambda},~~\widetilde{t}=\frac{c_2}{\lambda}\,t,\\
    \widetilde{\tau}_{mn}=\frac{\tau_{mn}}{\mu+\kappa},~~ \widetilde{\sigma}_{mn}=\frac{\sigma_{mn}}{\mu+\kappa}, ~~\widetilde{\mathfrak{M}}_{mn}=\frac{\mathfrak{M}_{mn}}{\left(\mu+\kappa\right)\lambda}, ~~\widetilde{\Pi}_{mn}=\frac{\Pi_{mn}}{\left(\mu+\kappa\right)\lambda}\\
    J=\frac{j}{\lambda^2}, ~~\widetilde{\Phi}_2=\Phi_2,~~ \widetilde{u}_m=\frac{u_m}{\lambda},~m,n=1,2,3
\end{gather*}
Following the approach of Chebakov et al. \cite{Chebakov2017} and Kaplunov et al. \cite{Kaplunov2022}, we decompose the non-local force stresses, and couple stresses into slow and fast components as, 
\begin{eqnarray}
    &&\text{for non-local force stresses:}~~~
    \renewcommand{\arraystretch}{1.1}
    \left. \begin{array}{lr}
        \widetilde{\tau}_{11}=p_{11}+q_{11}\\
        \widetilde{\tau}_{31}=p_{31}+\epsilon\,q_{31}\\
        \widetilde{\tau}_{13}=p_{13}+\epsilon\,q_{31}\\
        \widetilde{\tau}_{33}=p_{33}+\epsilon^2\, q_{33}
    \end{array}
	\right\} \label{32}\\
    &&\text{for non-local couple stresses:}~~~
    \renewcommand{\arraystretch}{1.1}
    \left. \begin{array}{lr}
        \widetilde{\mathfrak{M}}_{12}=r_{12}+\epsilon\, s_{12}\\
        \widetilde{\mathfrak{M}}_{32}=r_{32}+\epsilon^2\, s_{32}
    \end{array}
	\right\} \label{33}
\end{eqnarray}
where $p_{mn}, r_{mn}$ for $m,n=1,2,3$ represents the slow-varying components of the force stresses and couple stresses, respectively, while $q_{mn}, s_{mn}$ for $m,n=1,2,3$ represents the fast-varying components of force stresses and couple stresses, respectively. In the context of a micropolar solid, it is significant to acknowledge that the force stresses lose their symmetry. Specifically, for slow-varying quantities, we assume that the quantitites $p_{mn} \neq p_{nm}$. However, in contrast, the fast-varying stress components within the boundary layer retain symmetry, i.e $q_{mn}=q_{nm}$. This symmetry vanishes as we move away from the boundary layer.\\
As a result, the governing equations are reformulated into slow and fast-varying components, yielding the following expressions,
 \begin{eqnarray}
    &&\hspace{1cm}\renewcommand{\arraystretch}{1.6}
    \left. \begin{array}{lr}
         \frac{\partial p_{1n}}{\partial \chi}+\frac{\partial p_{3n}}{\partial \eta_s}=\frac{\partial^2 \widetilde{u}_n}{\partial \tilde{t}^2},~~
         \frac{\partial q_{1m}}{\partial \chi}+\frac{\partial q_{3m}}{\partial \eta_f}=0\\
         \frac{\partial r_{12}}{\partial \chi}+\frac{\partial r_{32}}{\partial \eta_s}+p_{31}-p_{13}=J\,\frac{\partial^2 \widetilde{\Phi}_2}{\partial \widetilde{t}^2}, ~~\frac{\partial s_{12}}{\partial \chi}+\frac{\partial s_{32}}{\partial \eta_f}=0
    \end{array}
    \right\} \label{34}\\
    &&\text{and}~~~
    \renewcommand{\arraystretch}{1.1}
    \left. \begin{array}{lr}
         \epsilon^2 \left(\frac{\partial p_{mn}}{\partial \chi^2}+\frac{\partial^2 p_{mn}}{\partial \eta_s^2}\right)-p_{mn}=-\widetilde{\sigma}_{mn}, ~~ \epsilon^2 \frac{\partial^2 q_{mn}}{\partial \chi^2}+ \frac{\partial^2 q_{mn}}{\partial \eta_f^2}-q_{mn}=0\\
         \epsilon^2 \left(\frac{\partial r_{m2}}{\partial \chi^2}+\frac{\partial^2 r_{mn}}{\partial \eta_s^2}\right)-r_{mn}=-\widetilde{\Pi}_{mn}, ~~ \epsilon^2 \frac{\partial^2 s_{mn}}{\partial \chi^2}+ \frac{\partial^2 s_{mn}}{\partial \eta_f^2}-s_{mn}=0
    \end{array}
    \right\} \label{35}\\
    &&\text{with}~~~
    \renewcommand{\arraystretch}{1.4}
    \left. \begin{array}{lr}
        \widetilde{\sigma}_{11}=\alpha_1^2\,\frac{\partial \widetilde{u}_1}{\partial \chi}+\left(\alpha_1^2-2+\alpha_2^2\right)\frac{\partial \widetilde{u}_3}{\partial \eta_s}\\
        \widetilde{\sigma}_{13}=\frac{\partial \widetilde{u}_3}{\partial \chi}+\left(1-\alpha_2^2\right)\frac{\partial \widetilde{u}_1}{\partial \eta_s}+\alpha_2^2\,\widetilde{\Phi}_2\\
        \widetilde{\sigma}_{31}=\frac{\partial \widetilde{u}_1}{\partial \eta_s}+\left(1-\alpha_2^2\right)\frac{\partial \widetilde{u}_3}{\partial \chi}-\alpha_2^2\,\widetilde{\Phi}_2\\
        \widetilde{\sigma}_{33}=\alpha_1^2\,\frac{\partial \widetilde{u}_3}{\partial \eta_s}+\left(\alpha_1^2-2+\alpha_2^2\right)\frac{\partial \widetilde{u}_1}{\partial \chi}\\
        \widetilde{\Pi}_{12}=\alpha_3^2\,J\, \frac{\partial \widetilde{\Phi}_2}{\partial \chi}\\
        \widetilde{\Pi}_{32}=\alpha_3^2\,J\, \frac{\partial \widetilde{\Phi}_2}{\partial \eta_s}       
         \end{array}
   		 \right\} \label{36}
 \end{eqnarray}
 subjected to the boundary conditions,
 \begin{gather}
 	 \renewcommand{\arraystretch}{1.6}
 	\left. 
 	\begin{array}{lr}
			p_{31}\Bigr|_{\eta_s=0}+\epsilon\,q_{31}\Bigr|_{\eta_f=0}=0,~~~\\
			p_{33}\Bigr|_{\eta_s=0}+\epsilon^2\,q_{33}\Bigr|_{\eta_f=0}=0,~~~\\
			r_{32}\Bigr|_{\eta_s=0}+\epsilon^2\,s_{32}\Bigr|_{\eta_f=0}=0\\
			p_{11}\Bigr|_{\eta_s=0}+q_{11}\Bigr|_{\eta_f=0}-\epsilon\,\frac{\partial p_{11}}{\partial \eta_s}\Bigr|_{\eta_s=0}-\frac{\partial q_{11}}{\partial \eta_f}\Bigr|_{\eta_f=0}-\frac{\epsilon^2}{2}\,\frac{\partial^3 q_{11}}{\partial \chi^2\,\partial \eta_f}\Bigr|_{\eta_f=0}=0\\
			r_{12}\Bigr|_{\eta_s=0}+\epsilon\,s_{12}\Bigr|_{\eta_f=0}-\epsilon\, \frac{\partial r_{12}}{\partial \eta_s}\Bigr|_{\eta_s=0}-\epsilon\,\frac{\partial s_{12}}{\partial \eta_f}\Bigr|_{\eta_f=0}=0 
	\end{array}
	\right\} \label{37}
 \end{gather}
 where $\alpha_1=\frac{c_1}{c_2},~\alpha_2=\frac{c_3}{c_2},~\alpha_3=\frac{c_4}{c_2}$ are the dimensionless quantities.\\
 The quantities $p_{mn}, q_{mn}, \widetilde{\sigma}_{mn}, r_{mn}, s_{mn}, \widetilde{\Pi}_{mn}, \widetilde{u}_{n}, \widetilde{\Phi}_2$ are now expanded in an asymptotic series of the form,
 \begin{eqnarray}
 	g=g^{(0)}+\epsilon\, g^{(1)}+\epsilon^2\,g^{(2)}+\dots \label{38}
 \end{eqnarray}
 where $g\in\{p_{mn}, q_{mn}, \widetilde{\sigma}_{mn}, r_{mn}, s_{mn}, \widetilde{\Pi}_{mn}, \widetilde{u}_{n}, \widetilde{\Phi}_2\}$.\\
 As a result of the asymptotic expansion sought in Eq. \eqref{38}, we can rewrite the Eqs. \eqref{34}$-$\eqref{37} for different asymptotic orders $i=0,1,2,\dots,$ as,
  \begin{eqnarray}
 	&&\hspace{1cm}\renewcommand{\arraystretch}{1.6}
 	\left. \begin{array}{lr}
 		\frac{\partial p_{1n}^{(i)}}{\partial \chi}+\frac{\partial p_{3n}^{(i)}}{\partial \eta_s}=\frac{\partial^2 \widetilde{u}_n^{(i)}}{\partial \tilde{t}^2},~~
 		\frac{\partial q_{1m}^{(i)}}{\partial \chi}+\frac{\partial q_{3m}^{(i)}}{\partial \eta_f}=0\\
 		\frac{\partial r_{12}^{(i)}}{\partial \chi}+\frac{\partial r_{32}^{(i)}}{\partial \eta_s}+p_{31}^{(i)}-p_{13}^{(i)}=J\,\frac{\partial^2 \widetilde{\Phi}_2^{(i)}}{\partial \widetilde{t}^2}, ~~\frac{\partial s_{12}^{(i)}}{\partial \chi}+\frac{\partial s_{32}^{(i)}}{\partial \eta_f}=0
 	\end{array}
 	\right\} \label{39}\\
 	&&\text{and}~~~
 	\renewcommand{\arraystretch}{1.1}
 	\left. \begin{array}{lr}
 		\frac{\partial^2 p_{mn}^{(i-2)}}{\partial \chi^2}+\frac{\partial^2 p_{mn}^{(i-2)}}{\partial \eta_s^2}-p_{mn}^{(i)}=-\widetilde{\sigma}_{mn}^{(i)}, ~~  \frac{\partial^2 q_{mn}^{(i-2)}}{\partial \chi^2}+ \frac{\partial^2 q_{mn}^{(i)}}{\partial \eta_f^2}-q_{mn}^{(i)}=0\\
 		 \frac{\partial^2 r_{m2}^{(i-2)}}{\partial \chi^2}+\frac{\partial^2 r_{mn}^{(i-2)}}{\partial \eta_s^2}-r_{mn}^{(i)}=-\widetilde{\Pi}_{mn}^{(i)}, ~~  \frac{\partial^2 s_{mn}^{(i-2)}}{\partial \chi^2}+ \frac{\partial^2 s_{mn}^{(i)}}{\partial \eta_f^2}-s_{mn}^{(i)}=0
 	\end{array}
 	\right\} \label{40}\\
 	&&\text{with}~~~
 	\renewcommand{\arraystretch}{1.6}
 	\left. \begin{array}{lr}
 		\widetilde{\sigma}_{11}^{(i)}=\alpha_1^2\,\frac{\partial \widetilde{u}_1^{(i)}}{\partial \chi}+\left(\alpha_1^2-2+\alpha_2^2\right)\frac{\partial \widetilde{u}_3^{(i)}}{\partial \eta_s}\\
 		\widetilde{\sigma}_{13}^{(i)}=\frac{\partial \widetilde{u}_3^{(i)}}{\partial \chi}+\left(1-\alpha_2^2\right)\frac{\partial \widetilde{u}_1^{(i)}}{\partial \eta_s}+\alpha_2^2\,\widetilde{\Phi}_2^{(i)}\\
 		\widetilde{\sigma}_{31}^{(i)}=\frac{\partial \widetilde{u}_1^{(i)}}{\partial \eta_s}+\left(1-\alpha_2^2\right)\frac{\partial \widetilde{u}_3^{(i)}}{\partial \chi}-\alpha_2^2\,\widetilde{\Phi}_2^{(i)}\\
 		\widetilde{\sigma}_{33}^{(i)}=\alpha_1^2\,\frac{\partial \widetilde{u}_3^{(i)}}{\partial \eta_s}+\left(\alpha_1^2-2+\alpha_2^2\right)\frac{\partial \widetilde{u}_1^{(i)}}{\partial \chi}\\
 		\widetilde{\Pi}_{12}^{(i)}=\alpha_3^2\,J\, \frac{\partial \widetilde{\Phi}_2^{(i)}}{\partial \chi}\\
 		\widetilde{\Pi}_{32}^{(i)}=\alpha_3^2\,J\, \frac{\partial \widetilde{\Phi}_2^{(i)}}{\partial \eta_s}       
 	\end{array}
 	\right\} \label{41}
 \end{eqnarray}
and the boundary conditions at the surface $\eta_s=\eta_f=0$ as,
 \begin{gather}
 	\renewcommand{\arraystretch}{1.6}
 	\left. 
 	\begin{array}{lr}
 		p_{31}^{(i)}\Bigr|_{\eta_s=0}+q_{31}^{(i-1)}\Bigr|_{\eta_f=0}=0,\\
 		p_{33}^{(i)}\Bigr|_{\eta_s=0}+q_{33}^{(i-2)}\Bigr|_{\eta_f=0}=0,\\
 		r_{32}^{(i)}\Bigr|_{\eta_s=0}+s_{32}^{(i-2)}\Bigr|_{\eta_f=0}=0\\
 		p_{11}^{(i)}\Bigr|_{\eta_s=0}+q_{11}^{(i)}\Bigr|_{\eta_f=0}-\frac{\partial p_{11}^{(i-1)}}{\partial \eta_s}\Bigr|_{\eta_s=0}-\frac{\partial q_{11}^{(i)}}{\partial \eta_f}\Bigr|_{\eta_f=0}-\frac{1}{2}\,\frac{\partial^3 q_{11}^{(i-2)}}{\partial \chi^2\,\partial \eta_f}\Bigr|_{\eta_f=0}=0\\
 		r_{12}^{(i)}\Bigr|_{\eta_s=0}+s_{12}^{(i-1)}\Bigr|_{\eta_f=0}- \frac{\partial r_{12}^{(i-1)}}{\partial \eta_s}\Bigr|_{\eta_s=0}- \frac{\partial s_{12}^{(i-1)}}{\partial \eta_f}\Bigr|_{\eta_f=0}=0
 	\end{array}
 	\right\} \label{42}
 \end{gather}
 Comparing the leading order terms in Eqs. \eqref{40}, we have
 \begin{gather}
     p_{mn}^{(0)}=\widetilde{\sigma}_{mn}^{(0)}, ~~~
     r_{mn}^{(0)}=\widetilde{\Pi}_{mn}^{(0)} \label{43}\\
    \frac{\partial^2 q_{mn}^{(0)}}{\partial \eta_f^2}-q_{mn}^{(0)}=0, ~~~
    \frac{\partial^2 s_{mn}^{(0)}}{\partial \eta_f^2}-s_{mn}^{(0)}=0 \label{44}
 \end{gather}
 The differential equations in Eqs. \eqref{44} are solved for the decaying solution to obtain,
 \begin{equation}
     q_{mn}^{(0)}=Q_{mn}^{(0)}\left(\chi, \tilde{t}\right)\, e^{-\eta_f}, ~~~s_{mn}^{(0)}=S_{mn}^{(0)}\left(\chi, \tilde{t}\right)\, e^{-\eta_f} \label{45}
 \end{equation}
where the values for some of the coefficients $Q_{mn}^{(0)}$ can be easily obtained using Eqs. \eqref{42}(iv) and \eqref{39} as,
 \begin{equation}
    Q_{11}^{(0)}=-\frac{1}{2}\, \widetilde{\sigma}_{11}^{(0)}\Bigr|_{\eta_s=0},~~~
    Q_{31}^{(0)}=-\frac{1}{2}\, \frac{\partial \widetilde{\sigma}_{11}^{(0)}}{\partial \chi}\Biggr|_{\eta_s=0},~~~
    Q_{33}^{(0)}=-\frac{1}{2}\, \frac{\partial^2 \widetilde{\sigma}_{11}^{(0)}}{\partial \chi^2}\Biggr|_{\eta_s=0} \label{46}
 \end{equation}
However, the values of $S_{mn}^{(0)}$ cannot be evaluated from the leading order term analysis.\\
As a result, the boundary conditions at the leading order becomes,
\begin{equation}
    \widetilde{\sigma}_{31}^{(0)}=0,~~~\widetilde{\sigma}_{33}^{(0)}=0,~~~ \widetilde{\Pi}_{32}^{(0)}=0 \label{47}
\end{equation}
which is the classical boundary conditions in a local micropolar elastic half-space.\\
Further analysing the first order terms at $i=1$ for Eq. \eqref{40}, we get
\begin{gather}
    p_{mn}^{(1)}=\widetilde{\sigma}_{mn}^{(1)}, ~~~r_{mn}^{(1)}=\widetilde{\Pi}_{mn}^{(1)}, \label{48}\\
    \frac{\partial^2 q_{mn}^{(1)}}{\partial \eta_f^2}-q_{mn}^{(1)}=0. \label{49}
\end{gather}
The solution to the differential equation \eqref{49} takes the form,
\begin{equation}
     q_{mn}^{(1)}=Q_{mn}^{(1)}\left(\chi, \tilde{t}\right)\, e^{-\eta_f}, \label{50}
 \end{equation}
 where
 \begin{gather}
    \renewcommand{\arraystretch}{1.6}
 	\left. 
 	\begin{array}{ll}
         Q_{11}^{(1)}=-\frac{1}{2}\, \left(\widetilde{\sigma}_{11}^{(1)}\Bigr|_{\eta_s=0}-\frac{\partial \widetilde{\sigma}_{11}^{(0)}}{\partial \eta_s}\Bigr|_{\eta_s=0}\right)\\
        Q_{31}^{(1)}=-\frac{1}{2}\, \left(\frac{\partial \widetilde{\sigma}_{11}^{(1)}}{\partial \chi}\Bigr|_{\eta_s=0}-\frac{\partial^2 \widetilde{\sigma}_{11}^{(0)}}{\partial \eta_s\,\partial \chi}\Bigr|_{\eta_s=0}\right)\\
        Q_{33}^{(1)}=-\frac{1}{2}\, \left(\frac{\partial ^2\widetilde{\sigma}_{11}^{(1)}}{\partial \chi^2}\Bigr|_{\eta_s=0}-\frac{\partial^3 \widetilde{\sigma}_{11}^{(0)}}{\partial \eta_s\,\partial \chi^2}\Bigr|_{\eta_s=0}\right)
    \end{array}
 	\right\} \label{51}     
 \end{gather}
 Moreover, using the boundary conditions in Eq. \eqref{42}, the values of $S_{mn}^{(0)}$ is obtained as,
 \begin{gather}
    \renewcommand{\arraystretch}{1.6}
 	\left. 
 	\begin{array}{ll}
        S_{12}^{(0)}=-\frac{1}{2}\, \left(\widetilde{\Pi}_{12}^{(1)}\Bigr|_{\eta_s=0}-\frac{\partial \widetilde{\Pi}_{12}^{(0)}}{\partial \eta_s}\Bigr|_{\eta_s=0}\right)\\
        S_{32}^{(0)}=-\frac{1}{2}\,\left(
        \frac{\partial \widetilde{\Pi}_{12}^{(1)}}{\partial \chi}\Bigr|_{\eta_s=0}-\frac{\partial^2\widetilde{\Pi}_{12}^{(0)}}{\partial \eta_s\,\partial \chi}\Bigr|_{\eta_s=0}\right)
    \end{array}
 	\right\} \label{52}
 \end{gather}
 Subsequently, the boundary conditions at the first order results in,
 \begin{eqnarray}
 	\widetilde{\sigma}_{31}^{(1)}-\frac{1}{2}\, \frac{\partial \widetilde{\sigma}_{11}^{(0)}}{\partial \chi}=0,~~~\widetilde{\sigma}_{33}^{(1)}=0, ~~~\widetilde{\Pi}_{32}^{(1)}=0 \label{53}
 \end{eqnarray}
 It is noteworthy that in the first-order boundary conditions of a non-local half-space, an additional term dependent on $\widetilde{\sigma}_{11}$ appears.\\
 Now, at the second order, the Eqs. \eqref{40} can be expressed using Eq. \eqref{43} as
 \begin{eqnarray}
    \renewcommand{\arraystretch}{1.6}
 	\left. 
 	\begin{array}{ll}
         \widetilde{\sigma}_{mn}^{(2)}+\frac{\partial^2 \widetilde{\sigma}_{mn}^{(0)}}{\partial \chi^2}+\frac{\partial^2 \widetilde{\sigma}_{mn}^{(0)}}{\partial \eta_s^2}=p_{mn}^{(2)}\\
         \widetilde{\Pi}_{32}^{(2)}+\frac{\partial^2 \widetilde{\Pi}_{32}^{(0)}}{\partial \chi^2}+\frac{\partial^2 \widetilde{\Pi}_{32}^{(0)}}{\partial \eta_s^2}=r_{32}^{(2)}  
     \end{array}
 	\right\} \label{54}
 \end{eqnarray}
This rewrites the equation of motion presented in Eq. \eqref{39} as,
 \begin{gather}
    \renewcommand{\arraystretch}{1.6}
 	\left. 
 	\begin{array}{ll}
         \frac{\partial \widetilde{\sigma}_{1n}^{(2)}}{\partial \chi}+\frac{\partial \widetilde{\sigma}_{3n}^{(2)}}{\partial \eta_s}=\frac{\partial^2 \widetilde{u}_n^{(2)}}{\partial \tilde{t}^2}-\frac{\partial^2}{\partial \tilde{t}^2}\left(\frac{\partial^2 \widetilde{u}_n^{(0)}}{\partial \chi^2}+\frac{\partial^2 \widetilde{u}_n^{(0)}}{\partial \eta_s^2}\right),\\
         \frac{\partial \widetilde{\Pi}_{12}^{(2)}}{\partial \chi}+\frac{\partial \widetilde{\Pi}_{32}^{(2)}}{\partial \eta_s}+\widetilde{\sigma}_{31}^{(2)}-\widetilde{\sigma}_{13}^{(2)}=J\frac{\partial^2 \tilde{\Phi}_2^{(2)}}{\partial \tilde{t}^2}-J \frac{\partial^2}{\partial \tilde{t}^2}\left(\frac{\partial^2 \tilde{\Phi}_2^{(0)}}{\partial \chi^2}+\frac{\partial^2 \tilde{\Phi}_2^{(0)}}{\partial \eta_s^2}\right). 
     \end{array}
 	\right\} \label{55}
 \end{gather}
 As a result, the boundary conditions at the second order can be written using Eq. \eqref{42} as,
 \begin{gather}
    \renewcommand{\arraystretch}{1.6}
 	\left. 
 	\begin{array}{ll}
        \widetilde{\sigma}_{31}^{(2)}+\frac{\partial^2 \widetilde{\sigma}_{31}^{(0)}}{\partial \chi^2}+\frac{\partial^2 \widetilde{\sigma}_{31}^{(0)}}{\partial \eta_s^2}-\frac{1}{2}\, \left(\frac{\partial \widetilde{\sigma}_{11}^{(1)}}{\partial \chi}-\frac{\partial^2 \widetilde{\sigma}_{11}^{(0)}}{\partial \eta_s\,\partial \chi}\right)=0\\
        \widetilde{\sigma}_{33}^{(2)}+\frac{\partial^2 \widetilde{\sigma}_{33}^{(0)}}{\partial \chi^2}+\frac{\partial^2 \widetilde{\sigma}_{33}^{(0)}}{\partial \eta_s^2}-\frac{1}{2}\, \frac{\partial^2 \widetilde{\sigma}_{11}^{(0)}}{\partial \chi^2}=0\\
        \widetilde{\Pi}_{32}^{(2)}+\frac{\partial^2 \widetilde{\Pi}_{32}^{(0)}}{\partial \chi^2}+\frac{\partial^2 \widetilde{\Pi}_{32}^{(0)}}{\partial \eta_s^2}-\frac{1}{2}\,\left(
        \frac{\partial \widetilde{\Pi}_{12}^{(1)}}{\partial \chi}-\frac{\partial^2\widetilde{\Pi}_{12}^{(0)}}{\partial \eta_s\,\partial \chi}\right)=0  
    \end{array}
 	\right\}\label{56}
\end{gather}
 Using the fact that, $f^{(1)}\approx\frac{1}{\epsilon}\,f,~ f^{(2)}\approx\frac{1}{\epsilon^2}\,f$ for $f\in\{\widetilde{\sigma}_{mn},\widetilde{\Pi}_{mn}\}$, the boundary value problem can now be given as,
 \begin{gather}
  \renewcommand{\arraystretch}{1.6}
 	\left. 
 	\begin{array}{ll}
     \frac{\partial \widetilde{\sigma}_{1n}}{\partial \chi}+\frac{\partial \widetilde{\sigma}_{3n}}{\partial \eta_s}=\frac{\partial^2 \widetilde{u}_n}{\partial \tilde{t}^2}-\epsilon^2\frac{\partial^2}{\partial \tilde{t}^2}\left(\frac{\partial^2 \widetilde{u}_n}{\partial \chi^2}+\frac{\partial^2 \widetilde{u}_n}{\partial \eta_s^2}\right),\\
     \frac{\partial \widetilde{\Pi}_{12}}{\partial \chi}+\frac{\partial \widetilde{\Pi}_{32}}{\partial \eta_s}+\widetilde{\sigma}_{31}-\widetilde{\sigma}_{13}=J\frac{\partial^2 \tilde{\Phi}_2}{\partial \tilde{t}^2}-J \epsilon^2\frac{\partial^2}{\partial \tilde{t}^2}\left(\frac{\partial^2 \tilde{\Phi}_2}{\partial \chi^2}+\frac{\partial^2 \tilde{\Phi}_2}{\partial \eta_s^2}\right),
      \end{array}
 	\right\}\label{57}
 \end{gather}
subjected to,
\begin{gather}
 \renewcommand{\arraystretch}{1.6}
 	\left. 
 	\begin{array}{ll}
    \widetilde{\sigma}_{31} -\frac{\epsilon}{2}\frac{\partial \widetilde{\sigma}_{11}}{\partial \chi}+\epsilon^2\left(\frac{\partial^2 \widetilde{\sigma}_{31}}{\partial \chi^2}+\frac{\partial^2 \widetilde{\sigma}_{31}}{\partial \eta_s^2}+\frac{1}{2}\,\frac{\partial^2 \widetilde{\sigma}_{11}}{\partial \eta_s\,\partial \chi}\right)=0\\
    \widetilde{\sigma}_{33}+\epsilon^2\left(\frac{\partial^2 \widetilde{\sigma}_{33}}{\partial \chi^2}+\frac{\partial^2 \widetilde{\sigma}_{33}}{\partial \eta_s^2}-\frac{1}{2}\, \frac{\partial^2 \widetilde{\sigma}_{11}}{\partial \chi^2}\right)=0\\
    \widetilde{\Pi}_{32}-\frac{\epsilon}{2}\frac{\partial \widetilde{\Pi}_{12}}{\partial \chi}+\epsilon^2\left(\frac{\partial^2 \widetilde{\Pi}_{32}}{\partial \chi^2}+\frac{\partial^2 \widetilde{\Pi}_{32}}{\partial \eta_s^2}+\frac{1}{2}\,\frac{\partial^2\widetilde{\Pi}_{12}}{\partial \eta_s\,\partial \chi}\right)=0 
     \end{array}
 	\right\}\label{58}
\end{gather}
 \section{Refined boundary value problem}
After recasting the refined boundary conditions in terms of the original variables employed throughout the analysis, the governing equations describing Rayleigh wave propagation on the surface of a non-local micropolar half-space can be expressed as:
\begin{eqnarray}
    \renewcommand{\arraystretch}{1.6}
    \left.
    \begin{array}{ll}
         \frac{\partial \sigma_{11}}{\partial x}+\frac{\partial \sigma_{31}}{\partial z}=\rho\left(1-\mathfrak{a}^2\nabla^2\right)\frac{\partial^2 u_1}{\partial t^2},\\
         \frac{\partial \sigma_{13}}{\partial x}+\frac{\partial \sigma_{33}}{\partial z}=\rho\left(1-\mathfrak{a}^2\nabla^2\right)\frac{\partial^2 u_3}{\partial t^2},\\
         \frac{\partial \Pi_{12}}{\partial x}+\frac{\partial \Pi_{32}}{\partial z}+\sigma_{31}-\sigma_{13}=\rho\,j\left(1-\mathfrak{a}^2\nabla^2\right)\frac{\partial^2 \tilde{\Phi}_2}{\partial t^2},
    \end{array}
 	\right\} \label{59}
\end{eqnarray}
 with the refined boundary conditions prescribed at the surface $z=0$ as,
 \begin{gather}
    \renewcommand{\arraystretch}{1.6}
 	\left. 
 	\begin{array}{ll}
        \sigma_{31} -\frac{\mathfrak{a}}{2}\frac{\partial \sigma_{11}}{\partial x}+\mathfrak{a}^2\left(\frac{\partial^2 \sigma_{31}}{\partial x^2}+\frac{\partial^2 \sigma_{31}}{\partial z^2}+\frac{1}{2}\,\frac{\partial^2 \sigma_{11}}{\partial x\,\partial z}\right)=0\\
        \sigma_{33}+\mathfrak{a}^2\left(\frac{\partial^2 \sigma_{33}}{\partial x^2}+\frac{\partial^2 \sigma_{33}}{\partial z^2}-\frac{1}{2}\, \frac{\partial^2 \sigma_{11}}{\partial x^2}\right)=0\\
        \Pi_{32}-\frac{\mathfrak{a}}{2}\frac{\partial \Pi_{12}}{\partial x}+\mathfrak{a}^2\left(\frac{\partial^2 \Pi_{32}}{\partial x^2}+\frac{\partial^2 \Pi_{32}}{\partial z^2}+\frac{1}{2}\,\frac{\partial^2 \Pi_{12}}{\partial x\,\partial z}\right)=0 
    \end{array}
 	\right\} \label{60}
 \end{gather}

\section{Conclusions}
    \par The equivalence of non-local integral and differential formulations breaks down for Rayleigh wave solutions in a non-local micropolar half-space. The solution derived using the differential formulation, which incorporates integral representations of non-local force and couple stress-free boundary conditions, exhibits inconsistencies with the original non-local equations of motion. This study adopts a similar approach as in Kaplunov et al. \cite{Kaplunov2022} by developing a modified singularly perturbed differential model for non-local micropolar elasticity. The model assumes a scenario where the characteristic wavelength significantly exceeds the internal material length scale. This analysis concludes that equivalence between the formulations is only achievable under specific additional boundary conditions at the surface. To ensure the well-posedness of the boundary value problem, the equivalence between the formulations is necessarily restricted to a specific set of force and couple stress distributions.
    \par The formation of a heterogeneous boundary layer necessitates a closer examination of the near-surface behavior in a non-local micropolar half-space. This boundary layer disrupts the homogeneity of the half-space in the vicinity of the surface, rendering Eringen's results on non-local micropolar elasticity inapplicable in this region.
    \par To address the influence of the boundary layer, an asymptotic analysis is employed using a small perturbation parameter. This analysis differentiates "slow" and "fast" variables along the length scale. Slow variables capture the slower overall dynamics of the system, while fast variables account for the rapid changes within the boundary layer.
    \par This investigation reveals that the impact of the boundary layer can be meticulously incorporated through the refinement of established traditional boundary conditions. This refinement manifests as second-order corrections to the conventional traction-free and couple stress-free conditions at the surface. These corrections account for the effects of other stresses arising from the additional boundary conditions and explicitly capture the impact of non-local phenomena near the micropolar surface.
 \subsection*{Declaration of competing interests}
 The authors affirm that they have no competing interests.
 \subsection*{Acknowledgements}
 The authors gratefully acknowledge the Applied Mathematics and Geomechanics (AMG) Lab, Indian Institute of Technology Indore, for providing the research facilities that made this work possible. Additionally, one of the authors acknowledges the financial support received through the PhD grant provided by the PMRF Scheme, Government of India, under ID number: 2101707.

 \end{document}